\documentclass[]{JHEP3}

\usepackage{graphicx}
\usepackage{graphics}
\usepackage{epsfig}
\epsfclipon

\usepackage{epsfig,bm}
\usepackage{latexsym}
\usepackage{amssymb,amsmath}

\newcommand{\nco}{\newcommand}
\nco{\beq}{\begin{equation}} \nco{\eeq}{\end{equation}}
\nco{\beqa}{\begin{eqnarray}} \nco{\eeqa}{\end{eqnarray}}

\def\be{\begin{equation}}
\def\ee{\end{equation}}

\def\baray{\begin{eqnarray}}
\def\earay{\end{eqnarray}}

\nco{\sss}{\scriptscriptstyle} \nco{\dphi}{\varphi}
\nco{\lsim}{\mbox{\raisebox{-.6ex}{~$\stackrel{<}{\sim}$~}}}
\nco{\gsim}{\mbox{\raisebox{-.6ex}{~$\stackrel{>}{\sim}$~}}}

\def\IK{\relax{\rm I\kern-.20em K}}
\def\IM{\relax{\rm I\kern-.20em M}}

\def\lsim{\mbox{\raisebox{-.6ex}{~$\stackrel{<}{\sim}$~}}}
\def\gsim{\mbox{\raisebox{-.6ex}{~$\stackrel{>}{\sim}$~}}}
\def\sss{\scriptscriptstyle}

\def\grad{\vec{\nabla}}

\title{Dynamics with Infinitely Many Derivatives: Variable Coefficient Equations}

\author{Neil Barnaby \\ Canadian Institute for Theoretical Astrophysics,
University of Toronto, 60 St.\ George St.\, Toronto, Ontario M5S 3H8 Canada \\
Email: \email{barnaby@cita.utoronto.ca}}

\author{Niky Kamran \\ Department of Mathematics and Statistics, McGill University,
Montr\'eal, Qu\'ebec, H3A 2K6 Canada \\ Email: \email{nkamran@math.mcgill.ca}}

\preprint{} \abstract{  
Infinite order differential equations have come to play an increasingly significant role in theoretical physics.
Field theories with infinitely many derivatives are ubiquitous in string field theory and have attracted interest recently also from 
cosmologists.  
Crucial to any application is a firm understanding of the mathematical structure of infinite order
partial differential equations.  In our previous work we developed a formalism to study the initial value problem for linear infinite order 
equations with constant coefficients.  Our approach relied on the use of a contour integral representation for the functions under consideration.  In 
many applications, including the study of cosmological perturbations in nonlocal inflation, one must solve linearized partial differential 
equations about some time-dependent background.  This typically leads to variable coefficient equations, in which case the contour integral methods employed previously 
become inappropriate.  In this paper we develop the theory of a particular class of linear infinite order partial differential 
equations with variable coefficients.  Our formalism is particularly well suited to the types of equations that arise in nonlocal cosmological perturbation theory.
As an example to illustrate our formalism we compute the leading corrections to the scalar field perturbations in $p$-adic inflation and show explicitly that these 
are small on large scales.
}

\keywords{differential equations of infinite order, string field theory, $p$-adic strings, cosmology of theories beyond the SM}

\begin{document}           

\section{Introduction}
\label{intro_sec}

Applications of infinite order differential equations to theoretical physics have attracted considerable interest recently.  Such equations are ubiquitous
in string field theory \cite{SFT} (see \cite{SFT_rev} for a review) and also arise in a number of toy models of string theory such as the $p$-adic string \cite{p_adic,zwiebach} and discrete 
world-sheet models \cite{discrete}.  Moreover, such equations have recently attracted interest from cosmologists due to a wide array of novel cosmological behaviours \cite{phantom1}-\cite{nl_cosmo}.  
Of particular interest are nonlocal inflationary models \cite{pi}-\cite{mulryne}, such as $p$-adic inflation, which can provide a unique playground for studying string cosmology in an ultra-violet complete setting.  
Nonlocal inflation has the remarkable property that slow roll inflation can proceed even when the potential is naively too steep and may therefore offer a novel way to circumvent the difficulty 
of finding flat scalar field potentials in string theory.  This remarkable behaviour was first observed in \cite{pi} and subsequently generalized to a broader class of models in \cite{lidsey}
(see also \cite{ng1}).  The perturbative analysis of \cite{pi}-\cite{lidsey} was verified using fully nonlinear simulations in \cite{mulryne}.

Perhaps the most interesting feature of nonlocal inflation is the possibility of generating an observably large nongaussian contribution to the temperature anisotropies in the
cosmic microwave background.  In \cite{ng1} an estimate was provided for the nongaussianity in a wide class of nonlocal models and this estimate was verified by a more quantitative calculation
in \cite{ng2}.  It is worth noting that the prediction of \cite{ng1} is consistent with a subsequent claimed detection of nongaussianity in \cite{detect}.  The quantitative calculation of \cite{ng2} was made possible
by a special gauge choice and by working to only to zeroth order in the $\epsilon$ slow roll parameter, despite keeping up to first order in the $\eta$ parameter.  This is a consistent approach 
for the models under consideration due to the hierarchy $|\eta| \gg \epsilon$, which is typical in hill-top inflation models \cite{hill_top}.  However, more
generally one would like to be able to study cosmological perturbations in nonlocal inflation without making such assumptions.  Progress in this direction is stymied by the fact that the
nonlocal cosmological perturbation theory equations are extremely difficult to solve.  In \cite{niky} the theory of linear infinite order equations with constant coefficients was 
developed.\footnote{See also \cite{calcagni}-\cite{IVP} different approaches to the initial value problem, \cite{math} for mathematical analysis of $p$-adic and string field equations and \cite{nl_solns} for more details
on solving nonlocal equations.}  However, this theory relied on integral transform methods which fail to be useful for the variable coefficient equations that arise in cosmological perturbation theory.

To improve on the calculation of \cite{ng2}, then, clearly we require more sophisticated analytical tools and a better understanding of the formal aspects of infinite order equations.  In this paper 
we make progress in this direction by developing the theory of a broad class of infinite order variable coefficient equations.  Our formalism is particularly well suited to studying the 
kinds of equations that describe nonlocal cosmological perturbations.  This work is the first in a series of papers that aim to develop a general approach to nonlocal cosmological perturbation theory.  
As an illustration of our method, we will compute the leading corrections to the inflaton perturbations in $p$-adic inflation, showing explicitly that these are small and do not lead to super-horizon evolution.

It is appropriate, before moving on, to mention the difficulties and complications that arise when one wishes to interpret higher derivative theories as fundamental.  Such theories are often fraught with classical instabilities
known as Ostrogradski instabilities \cite{ostrogradski} (see \cite{1/r} for a modern review), as emphasized in \cite{woodard}.  
In this paper our primary interest is in developing the tools necessary to solve nonlocal
equations of motion in a very general context and we will not address the important question of when such theories can be phenomenologically viable.  See \cite{nl_cosmo,maximal} for examples of stable, interacting nonlocal theories.  
See \cite{mulryne_prog} for a discussion of the nonlinear stability of theories with infinitely many derivatives.

The organization of this paper is as follows.  In section \ref{eqns_sec} we introduce the class of equations under consideration and review previous results.  In section \ref{const_sec} we consider the
simplest possible extension of the analysis of \cite{niky}.  In section \ref{subst_sec} we give a formal solution of the class of equations under consideration which relies on knowledge of the
inverse of the nonlocal operator under consideration.  In section \ref{inverse_sec} we give two different methods for computing this inverse operator.
In section \ref{app_sec} we apply our methods to $p$-adic inflation.  Finally, in section \ref{concl_sec} we conclude by discussing further possible applications of our formalism.  In appendices
A and B we review the solutions for the inflationary background and scalar field perturbations in $p$-adic inflation.  In appendix C we discuss some technical manipulations involving Bessel 
functions.  
In appendix D we discuss the non-self-adjointness of the d'Alembertian in de Sitter space.

\section{Infinite Order Equations with Variable Coefficients}
\label{eqns_sec}

In this paper we develop the theory of a particular class of linear infinite order partial differential equations having variable coefficients.
The class of equation under consideration can be cast in the form
\begin{equation}
\label{proto}
  F(\Box)\phi(t,{\bf x}) = m^2(t)\phi(t,{\bf x})
\end{equation}
where $\Box = g^{\mu\nu}\nabla_{\mu}\nabla_{\nu}$ is the covariant d'Alembertian.\footnote{Our choice of metric signature is such that $g^{\mu\nu} \equiv \eta^{\mu\nu} = \mathrm{diag}(-1,+1,+1,+1)$
in flat space.}   Throughout we will assume that the kinetic function $F(z)$ (also called the \emph{generatrix} in the mathematics literature) is entire.  With this assumption $F(z)$ can be represented by a 
power series centered at $z=0$ and having infinite radius of convergence.  Thus, we define the action of the pseudo-differential operator $F(\Box)$ on some smooth function $\phi$ by the series expansion
\begin{equation}
\label{series_def}
  F(\Box)\phi \equiv \sum_{n=0}^{\infty} a_n \Box^n \phi
\end{equation}
where the powers $\Box^n$ are, of course, understood as composition of the operator $\Box$ with itself $n$ times and the coefficients in the expansion are
\begin{equation}
\label{a_n}
  a_n = \frac{F^{(n)}(0)}{n!}
\end{equation}
Our restriction to analytic $F(z)$ is merely for simplicity.  We expect that it should be straightforward to generalize our results to $F(z)$ having isolated poles
(such as the zeta strings model \cite{zeta}) or branch cuts (see \cite{niky} for mathematical analysis and \cite{root} for cosmological applications).

The motivation to consider (\ref{proto}) comes from studying a general class of infinite order equations 
\begin{equation}
\label{gen}
  F(\Box)\phi = V'\left[\phi\right]
\end{equation}
which are  typical in string field theory and also in cosmological models.  
Suppose one wishes to solve (\ref{gen}) for small inhomogeneities $\delta \phi(t,{\bf x})$ about some known homogeneous 
solution $\phi_0(t)$.  Plugging the ansatz
\begin{equation}
  \phi(t,{\bf x}) = \phi_0(t) + \delta \phi(t,{\bf x})
\end{equation}
into (\ref{gen}) and linearizing in $\delta \phi$ we obtain
\begin{equation}
  F(\Box)\delta \phi(t,{\bf x}) = V''\left[\phi_0(t)\right]\delta \phi(t,{\bf x})
\end{equation}
which is precisely of the form (\ref{proto}) with $m^2(t) \equiv V''\left[\phi_0(t)\right]$.

Equations of the form (\ref{proto}) with $m^2 = \mathrm{const}$ and $g^{\mu\nu} = \eta^{\mu\nu}$ were considered
in our previous work \cite{niky}.  In that paper solutions were derived using the formal operator calculus and we were able to exhaustively count the number of
initial data necessary to uniquely specify a solution.  The formalism developed in \cite{niky} relied on the fact that in flat space with $m^2 = \mathrm{const}$
(\ref{proto}) is an infinite order equation with constant coefficients.  However, in many applications both $g^{\mu\nu}$ and $m^2$ will depend nontrivially
on space-time coordinates.  In this case the contour integral approach adopted in \cite{niky} is no longer applicable.

A physical example where an equation of the form (\ref{proto}) arises is the computation of cosmological perturbations in nonlocal inflationary models.  In the case of $p$-adic inflation the dynamical
equation for the scalar field is
\begin{equation}
\label{p}
  p^{-\Box / (2m_s^2) }\phi = \phi^p
\end{equation}
For cosmological applications one should solve (\ref{p}) in an FRW geometry 
\begin{equation}
  ds^2\equiv g_{\mu\nu}dx^{\mu}dx^{\nu} = -dt^2 + a^2(t) dx_idx^i
\end{equation}
Slowly rolling background solutions $\phi_0(t)$ have been constructed which source a quasi-de Sitter expansion $H \equiv \dot{a} / a \cong \mathrm{const}$ in \cite{pi}.  Neglecting induced 
inhomogeneities of the metric, the cosmological perturbations satisfy the equation
\begin{equation}
\label{p_pert}
  \left[  p^{-\Box / (2m_s^2)} - p \right]\delta \phi(t,{\bf x}) = p \left[\phi_0^{p-1}(t) - 1 \right]\delta \phi(t,{\bf x})
\end{equation}
which is precisely of the form (\ref{proto}) with 
\begin{eqnarray}
  F(\Box) &=& p^{-\Box / (2m_s^2)} - p \label{pF} \\
  m^2(t) &=&  p \left[\phi_0^{p-1}(t) - 1 \right] \label{pm}
\end{eqnarray}

\section{The Case $m^2(t) = 0$}
\label{const_sec}

Let us first consider the simplest possible extension of the results of \cite{niky} by assuming that $m^2(t) \equiv 0$ but allowing for $g^{\mu\nu}$ to be nontrivial.\footnote{This analysis
applies equally well to $m^2(t) = \mathrm{const}$ since any additive constant could be absorbed into the definition of $F(\Box)$.}  Let us assume that the function $F(z)$ has $N$ zeroes, all of 
which are order unity (this is to be expected for physically interesting operators).  

To solve (\ref{proto}) we first provisionally assume that $\phi$ is a formal eigenfunction of $\Box$, meaning that it is a non-trivial solution of the equation
\begin{equation}
\label{evalue}
  \Box \phi = -\omega^2 \phi
\end{equation}
Acting on $\phi$ with the full operator $F(\Box)$ we have
\begin{eqnarray}
  F(\Box) \phi &=& \sum_{n=0}^{\infty}a_n \Box^n\phi \nonumber\\
                                    &=& \sum_{n=0}^{\infty}a_n (-\omega^2)^n \phi \nonumber\\
                                    &=& F(-\omega^2)\phi \nonumber
\end{eqnarray}
(The re-summation on the third line is justified for any value of $\omega$ because $F(z)$ is entire.)
Hence the solutions of (\ref{evalue}) will give rise to solutions of (\ref{proto}) provided the eigenvalue $\omega^2$ is chosen to satisfy 
the transcendental equation
\begin{equation}
\label{root}
  F(-\omega^2) = 0
\end{equation}
By assumption this equation has $N$ distinct roots $\omega_i^2$ ($i = 1,\cdots,N$) and hence there are $N$ distinct solutions of (\ref{proto}). We assume that to each of these roots, there corresponds a formal eigenfunction $\phi_i$ of the 
d'Alembertian.\footnote{There is of course a scaling ambiguity in the choice of $\phi_i$. In the context of the cosmological applications considered in this paper, the ambiguity will be resolved by considering solutions with prescribed 
asymptotics at infinity and at the horizon (that is simply the usual Bunch-Davies procedure, see Appendix B).}
The most general solution of (\ref{proto}) is obtained by superposing these eigenfunctions as
\begin{equation}
\label{const_full}
  \phi(t,{\bf x}) = \sum_{i=1}^N \phi_i(t,{\bf x})
\end{equation}
Each $\phi_i$ is a solution of the second order equation $\Box \phi_i = -\omega_i^2 \phi_i$ and hence contains two degrees of freedom.\footnote{By ``degrees of freedom''
here we refer to the freedom to specify two independent functions of the spatial variables $x_i$ to specify the Cauchy data corresponding to a solution.  This is equivalent to saying
that, upon quantization, each $\phi_i$ will have its own set of annihilation/creation operators.}  The full solution (\ref{const_full}) then admits
$2N$ initial data, two for each zero of $F(z)$.  This is precisely the same result that was obtained for flat space in \cite{niky} where a physical interpretation was provided 
in terms of the poles of the propagator. It should be remarked that we are not making any completeness assumption on the set of formal eigenfunctions. In fact, the construction being local, we need not even assume at this stage that the wave operator $\Box$ is essentially self-adjoint.

Let us briefly discuss the application of this approach to $p$-adic inflation.  Consider equation (\ref{p_pert}).  During inflation 
$\phi_0(t) \cong 1$ so that $m^2(t) \ll 1$ and (\ref{p_pert}) can be approximated by 
\begin{equation}
  \left[  p^{-\Box / (2 m_s^2) } - p \right] \delta \phi(t,{\bf x}) \cong 0
\end{equation}
which is of the form under consideration.  This equation is solved, as above, by taking $\delta \phi(t,{\bf x})$ to be an eigenfunction of $\Box$.  
This is precisely the approach that was adopted in \cite{pi} to study cosmological perturbations 
in $p$-adic inflation (and also in \cite{ng1,ng2} in a more general context).  In this paper we will to go beyond this approximation and show how to systematically include the effect of having $\phi_0(t)$ different from unity.

\section{The Method of Successive Substitution}
\label{subst_sec}

Let us now develop the theory of equation (\ref{proto}) in the case where the time dependence of $m^2(t)$ cannot be neglected.\footnote{In fact, the approach of this section may fail
in the case of constant $m^2(t)$.}   Let us suppose first that we are able to solve the simpler equation
\begin{equation}
\label{chi}
  F(\Box)\chi = 0
\end{equation}
This falls into the class of equations considered in section \ref{const_sec} and the solutions are eigenfunctions of $\Box$.  Denoting the $N$ solutions
of (\ref{chi}) by $\chi_i$ we have
\begin{eqnarray}
  \Box \chi_i &=& -\omega_i^2 \chi_i\\
  F(-\omega_i^2)&=& 0
\end{eqnarray}
for $i = 1, \cdots, N$.

Next, consider the inhomogeneous equation
\begin{equation}
\label{chiJ}
  F(\Box) \chi = J
\end{equation}
Let us suppose that we are able to solve this equation to obtain the particular solution $\chi_{\mathrm{par}}$.\footnote{We will discuss the solution
of this equation in more detail in the next section.}  Writing the particular solution as
\begin{equation}
\label{G}
  \chi_{\mathrm{par}} = G J
\end{equation}
defines the operator $G$, which is the inverse of the kinetic function $F(\Box)$.  In the mathematical literature
$G$ is often referred to as the \emph{resolvent generatrix}.

We now demonstrate that knowledge of the solutions of (\ref{chi}) and of the inverse operator $G$ is sufficient to construct the full solutions of (\ref{proto}).  
Consider the infinite series
\begin{eqnarray}
  \phi_i &=& \chi_i \nonumber \\
         &+& G\left[m^2 \chi_i \right] + G\left[m^2 G\left[ m^2 \chi_i \right ]\right] + G\left[m^2 G\left[ m^2 G\left[ m^2 \chi_i \right] \right ]\right] \nonumber \\
         &+& \cdots \label{phi_i}         
\end{eqnarray}
Plugging (\ref{phi_i}) into (\ref{proto}) and making use of the fact that $F(\Box)\chi_i = 0$ and that $G$ is the inverse of $F$ one can easily 
verify that $\phi_i$ affords a formal solution of (\ref{proto}).  The most general solution of (\ref{proto}) is obtained by superposing the modes $\phi_i$ as
\begin{equation}
  \phi(t,{\bf x}) = \sum_{i=1}^N \phi_i(t,{\bf x})
\end{equation}
Hence we conclude that $\phi(t,{\bf x})$ contains the same number of degrees of freedom as $\chi(t,{\bf x})$.  That is, $\phi(t,{\bf x})$ admits two initial data for each zero of $F(z)$.

The utility of the solution (\ref{phi_i}) lies in the observation that when $m^2(t)$ is a small perturbation the series of successive iterations on the second line takes the form
of an expansion in powers of a small parameter.\footnote{We will quantify what is meant by ``small'' on a case-by-case basis.}  In this case the series (\ref{phi_i}) 
can be truncated at some finite order to obtain an approximate expression for $\phi_i(t,{\bf x})$.  

Notice that the definition of the quantities $F(z)$ and $m^2(t)$ in (\ref{proto}) is ambiguous.  The form of this equation and also the expression for
the solution (\ref{phi_i}) is unchanged under the substitution $F(z) \rightarrow F(z) + A$, $m^2(t) \rightarrow m^2(t) + A$ for any constant $A$.  In applications,
we can take advantage of this ambiguity to define $m^2(t)$ in such a way that it is small and the series (\ref{phi_i}) can safely be truncated at low order.
Indeed, we have already implicitly done this in writing (\ref{p_pert}): by subtracting $p\delta\phi$ from both sides of the equation we have defined $F(\Box)$, $m^2(t)$
in such a way that $m^2(t) \ll 1$ during inflation.  With this definition the application of (\ref{phi_i}) to equation (\ref{p_pert}) has a nice physical interpretation.
The solution $\chi_i$ represent the pure de Sitter space modes constructed in \cite{pi} while the subsequent terms in the series (\ref{p_pert}) represent the leading order 
corrections that arise due to the slow motion of $\phi_0(t)$ away from the unstable maximum $\phi = 1$.

\section{The Inverse Operator}
\label{inverse_sec}

In the last section we showed that (\ref{phi_i}) affords a formal solution of (\ref{proto}).  This solution is given in terms of the solutions of the simpler equation 
(\ref{chi}) and the inverse operator, $G$.  In section \ref{const_sec} we showed that the construction of the solutions of (\ref{chi}) is straightforward once the 
eigenfunctions of $\Box$ are known.  Hence, the success of our method hinges on our ability to construct the  inverse operator $G$ associated with $F(\Box)$.  
In \cite{niky} we were able to obtain exact expressions of the solutions of the inhomogeneous equation (\ref{chiJ}) for the special case $g^{\mu\nu} = \eta^{\mu\nu}$
by using contour integral methods.  However, generalizing these results to curved backgrounds is a nontrivial task.  Below we discuss two approaches to computing 
the resolvent which will prove to be useful in applications.  It is worth emphasizing that the methods developed here are applicable to a broad class of inhomogeneous
nonlocal equations.

\subsection{Expansion of the Source in Eigenfunctions}
\label{J_efunctions_sec}

Let us first consider the case where $G$ acts on an eigenfunction of the d'Alembertian.  Hence, we wish to solve (\ref{chiJ}) when $J$ satisfies
\begin{equation}
\label{Jevalue}
  \Box J = m_J^2 J
\end{equation}
If we assume that $F(m_J^2) \not= 0$ then (\ref{chiJ}) is trivially solved by
\begin{equation}
  \chi_{\mathrm{par}} = \frac{1}{F(m_J^2)}J
\end{equation}
It follows that, when $G$ acts on an eigenfunction of $\Box$, we have the simple expression
\begin{equation}
\label{J_efunction_G}
  G(\Box) = \frac{1}{F(m_J^2)}
\end{equation}

Suppose, now, that $J$ is not an eigenfunction but it can be expanded into a sum of eigenfunctions as
\begin{equation}
\label{Jexpand}
  J = \sum_n c_n J_n
\end{equation}
where the $c_n$ are constant and $\Box J_n = m_n^2 J_n$.  Note that we are not assuming that the $J_n$ form a complete set of eigenfunctions of $\Box$. All we require at this stage is that $J$ should be 
expressible as a series in a set of formal or true eigenfunctions of $\Box$, which should be differentiable term by term infinitely many times.  Assuming that $F(m_n^2) \not= 0$ for all $n$ we have
\begin{equation}
\label{J_efunction_G_sum}
  G\left[J \right] =  \sum_n \frac{c_n}{F(m_n^2)} J_n
\end{equation}
owing to the linearity of the resolvent $G$.  We will see that the simple expression (\ref{J_efunction_G_sum}) 
will allow us to compute the corrections to the inflaton perturbations in $p$-adic inflation. 

Before moving on let us comment on the generality of this approach.  The form (\ref{Jexpand}) is quite special and we are not, in general, guaranteed that an arbitrary source $J(t,{\bf x})$
is expansible in eigenfunctions of $\Box$.  This is so because for nontrivial geometries the operator $\Box$ may fail to be self-adjoint when one imposes physically interesting boundary conditions on
the eigenfunctions.  For example, in de Sitter space the d'Alembertian is not self-adjoint when acting on a function space that includes Bunch-Davies normalized mode functions; see appendix D.
Of course, even when $\Box$ is not self-adjoint it may still happen that some particular physically interesting source term \emph{is} expansible in eigenfunctions, which is the case for
$p$-adic inflation.

\subsection{The Method of Infinite ``Differentiation''}
\label{method_sec}

As we have discussed, the case considered in the last subsection is rather special and an expansion of $J$ into eigenfunctions of $\Box$ may not always be possible.  
Thus, it may be interesting to develop another approach to computing the resolvent which applies more generally.  Here we develop an alternative formalism which is 
based on a modification of the ``method of infinite differentiation'' employed by Davis \cite{davis}.  The method developed in this subsection will not actually be
required for the explicit example we consider below and is included for completeness.

The method of infinite ``differentiation'' will furnish us with an expression for $G(\Box)$ as an infinite series
expansion in powers of $\Box$.  The price for the generality of our construction will be that the series cannot in general be summed to obtain a closed form
expression for $G$.
For reasons that will become clear shortly we assume that $F(0) \not= 0$ (so that $a_0 \not= 0$).\footnote{This does not entail any loss of generality since
if $F(0) = 0$ then we can redefine $F(z) \rightarrow A + F(z)$ and $m^2 \rightarrow A + m^2$ (where $A$ is any constant) without changing the form of (\ref{proto}).}  

Now, let us proceed with the construction of the resolvent.  Our aim is to solve the inhomogeneous equation (\ref{chiJ}).  To this end, consider the infinite series of equations 
formed by operating on (\ref{chiJ}) with successively higher powers of the operator $\Box$. That is, consider the series of equations 
\begin{eqnarray*}
  F(\Box) \chi &=& J \\
  \Box F(\Box) \chi &=& \Box J \\
\Box^2 F(\Box) \chi &=& \Box^2 J \\
 &\cdots&
\end{eqnarray*}
Writing $F(z)$ in terms of the expansion (\ref{series_def}) we then have the system of equations
\begin{equation} 
\label{eqns}
\begin{array}{ccccccc}
 a_0 \chi & + \,\,a_1 \Box \chi  & + \,\,a_2 \Box^2 \chi & +\,\, a_3 \Box^3 \chi & + \,\,\cdots & = & J   \\
 0  & + \,\, a_0\Box\chi  & +\, \,a_1 \Box^2 \chi & + \,\,a_2 \Box^3 \chi & +\,\, \cdots & = & \Box J   \\
 0  & + \,\, 0   & +\,\, a_0 \Box^2 \chi & +\,\, a_1 \Box^3 \chi & +\,\, \cdots & = & \Box^2 J \\
  \cdots  & \cdots   & \cdots & \cdots  & \cdots  & = & \cdots 
\end{array}
\end{equation}

One can now consider (\ref{eqns}) as an infinite system of equations in infinitely many unknowns $\chi$, $\Box \chi$, $\Box^2 \chi$, $\cdots$  This system can
be solved algebraically for the unknown $\chi$.  The equations (\ref{eqns}) can be considered as a matrix equation
\begin{equation}
\label{matrix}
  A\, V = S
\end{equation}
where the matrix $A$ is
\begin{equation}
\label{A}
  A = \left[ \begin{array}{ccccc}
a_0 & a_1 & a_2 & a_3 & \cdots \\
0 & a_0 & a_1 & a_2 & \cdots \\
0 & 0 & a_0 & a_1 & \cdots \\
0 & 0 & 0 & a_0 & \cdots \\
\cdots & \cdots & \cdots & \cdots & \cdots \end{array} \right]
\end{equation}
while $V$ is the vector of unknowns
\begin{equation}
\label{v}
  V = \left[ \begin{array}{c}
\chi  \\
\Box \chi \\
\Box^2 \chi \\
\Box^3 \chi \\
\cdots \end{array} \right]
\end{equation}
and the vector $S$ is constructed from the source
\begin{equation}
\label{S}
  S = \left[ \begin{array}{c}
J  \\
\Box J \\
\Box^2 J \\
\Box^3 J \\
\cdots \end{array} \right]
\end{equation}
Notice that the matrix $A$ is triangular (this is so because the coefficients $a_n$ defined by (\ref{a_n}) are constants).  This allows for a straightforward
iterative inversion of (\ref{eqns}).  Let us now illustrate how this works. As a first step we use the first line of (\ref{eqns}) to solve for $\chi$ in terms of $J$ and $\Box^n \chi$ with $n>1$:
\[
  \chi = \frac{1}{a_0}\left[ J -  a_1 \Box \chi  - a_2 \Box^2 \chi + \cdots \right]
\]
Next we use the second line of (\ref{eqns}) to eliminate $\Box \chi$ in favour of $\Box J$ and $\Box^n \chi$ with $n>2$, giving
\[
  \chi = \frac{1}{a_0}\left[ J -  \frac{a_1}{a_0} \Box J  + \left(\frac{a_1^2}{a_0} -  a_2\right) \Box^2 \chi + \cdots \right]
\]
Continuing in this matter we obtain an expression for $\phi$ solely in terms of $\Box^n J$:
\[
  \chi = \frac{1}{a_0}\left[ J - \frac{a_1}{a_0}\Box J  + \frac{1}{a_0^2}(a_1^2-  a_2a_0) \Box^2 J +  \cdots \right]
\]
Writing $\chi = G J$ we obtain a formal expression for the resolvent $G(\Box)$ as a power series expansion
\begin{equation}
  G(\Box) = \sum_{n=0}^{\infty} b_n \Box^n
\label{G2}
\end{equation}
where the explicit expressions for the first few coefficients are 
\begin{eqnarray}
  b_0 &=& \frac{1}{a_0} \\
  b_1 &=& -\frac{a_1}{a_0^2} \\
  b_2 &=& \frac{1}{a_0^3}(a_1^2-  a_0 a_2) \\
  b_3 &=& \frac{1}{a_0^4}(2a_0a_1a_2 - a_1^3 - a_0^2a_3)
\end{eqnarray}
At the formal level, this procedure can be continued to arbitrarily high order in $\Box^n$. The validity of the procedure will of course depend on the choice of source term $J$. 
This is in principle a potential limitation of the method. In fact, there are instances in which the expansion of $\chi$ in powers of $\Box J$ fails to converge pointwise. 

\section{Application to $p$-adic Inflation}
\label{app_sec}

\subsection{Setting Up the Calculation}

We now apply the formalism of sections \ref{subst_sec} and \ref{inverse_sec} to study the dynamics of scalar field perturbations during $p$-adic inflation, equations (\ref{p_pert}-\ref{pm}).
The homogeneous background solutions $\phi_0(t)$, $H(t)$ were constructed in \cite{pi} and these result are reviewed in appendix A.  The solutions are written as a series expansion in powers of the 
small parameter $u(t)$ defined by
\begin{equation}
  u(t) \equiv a(t)^{|\eta|} = e^{|\eta|H_0 t} \ll 1
\end{equation}
where
\begin{equation}
  \eta = -\frac{2m_s^2}{3H_0^2}
\end{equation}
is a slow roll parameter satisfying $|\eta| \ll 1$\footnote{In $p$-adic inflation the spectral index is given by $n_s - 1 \cong 2\eta$.
For the WMAP5-preferred value $n_s \cong 0.96$ \cite{WMAP5} we have $\eta \cong -0.02$.  We will use this value in our examples below.} and $H_0$ is the background
Hubble scale (written explicitly in terms of model parameters in appendix A).  To leading order the quantity $m^2(t)$ (defined explicitly in (\ref{pm})) is given by
\begin{equation}
  m^2(t) \cong -p(p-1)u(t)
\end{equation}
Because $m^2 \propto u$ it follows that the correction terms on the second line of (\ref{phi_i}) are controlled by the smallness of $u$.
If we work exclusively to leading order in $u$ we need only consider the first term on the second line of (\ref{phi_i}).

In what follows it will be simplest to work in terms of conformal time $\tau$ defined by $a d\tau = dt$ in terms of cosmic time $t$.  To leading order we have $a = e^{H_0 t} = -1/(H_0 \tau)$
so that $u(\tau) = (-H_0\tau)^{-|\eta|}$.  To the same accuracy the d'Alembertian takes the form
\begin{equation}
\label{box1}
  \Box = H_0^2 \left[-\tau^2 \partial_\tau^2 +2\tau\partial_\tau + \tau^2 \grad^2 \right]
\end{equation}

To proceed with the construction of the solution (\ref{phi_i}) we need two ingredients: the functions $\chi_i$ defined by (\ref{chi}) and an expression for the inverse operator $G$,
defined by (\ref{chiJ}).  Let us first discuss the solutions of (\ref{chi}).  The relevant functions $\chi_i$ were derived in \cite{pi} and these results are reviewed in appendix B.  
We consider only the tachyonic excitation with effective mass $-\omega_0^2 = -2m_s^2$ and thus drop the subscript $i$ on the solution.  Expanding $\chi$ in terms of annihilation/creation operators 
$a_{\bf k}$, $a_{\bf k}^{\dagger}$ and c-number valued modes functions $\chi_k$ we have
\begin{equation}
  \chi(t,{\bf x}) = \int \frac{d^3k}{(2\pi)^{3/2}}\left[a_{\bf k} \chi_k(\tau)e^{i{\bf k}\cdot {\bf x}} + \mathrm{h.c.}  \right]
\end{equation}
where $\mathrm{h.c.}$ denotes the Hermitian conjugate of the preceding term.  To leading order in $u$ and $\eta$ the mode functions can be written as
\begin{equation}
\label{mode}
  \chi_k(\tau) = \frac{H_{0}\sqrt{\pi}}{2k^{3/2}}(-k\tau)^{3/2}H_{3/2 - \eta}^{(1)}\left(-k\tau\right)
\end{equation}
up to an irrelevant constant phase.  For ease of notation we introduce the Fourier space d'Alembertian operator $\Box_k$ defined by
\begin{equation}
\label{box2}
  \Box_k = -H_0^2  \left[\tau^2 \partial_\tau^2 -2\tau\partial_\tau + (k \tau)^2 \right]
\end{equation}
By construction the function (\ref{mode}) obeys the eigenvalue equation
\begin{equation}
\label{chi_evalue}
  \Box_k \chi_k(\tau) = -2m_s^2\, \chi_k(\tau)
\end{equation}

Our goal is now to compute the corrections coming from the second line of (\ref{phi_i}).  To leading order in $u$ equation (\ref{phi_i})
takes the form
\begin{equation}
\label{large_corrections}
  \phi_k(\tau) - \chi_k(\tau) \cong G\left[ J_k(\tau)\right]
\end{equation}
where we have introduced the notation
\begin{eqnarray}
  J_k(\tau) &\equiv& m^2(\tau)\chi_k(\tau) \nonumber \\
            &=& -p(p-1) \frac{H_0\sqrt{\pi}}{2 k^{3/2}}\left(\frac{H_0}{k}\right)^{\eta} \left(-k\tau\right)^{3/2 + \eta} H_{3/2 - \eta}^{(1)}\left(-k\tau\right) \label{Jk}
\end{eqnarray}
Obviously to proceed further we will need an expression for the inverse operator $G$.  We could proceed by expanding $G$ as a power series
in $\Box$, following (\ref{G2}), and then simply computing $\Box_k^n J_k(\tau)$ to high order in $n$.  Unfortunately, this will turn out to give rather poor
convergence properties for the resulting series (this observation does \emph{not} imply that $G\left[J_k(\tau)\right]$ is divergent,
only that the approach of subsection \ref{method_sec} is not ideal for this particular problem).  The failure of the formalism from subsection \ref{method_sec} in 
this physically interesting case is disappointing, however, the simple trick discussed in subsection \ref{J_efunctions_sec} will be quite sufficient for our purposes.

\subsection{The Large Scale Limit}

As a warm-up exercise we first consider the limit $-k\tau \rightarrow 0$ which is relevant for cosmological observations.  In this limit the ``source'' term $J_k(\tau)$ given by eqn.\ (\ref{Jk})
obeys the eigenvalue equation
\begin{equation}
\label{Jk_evalue}
  \Box_k J_k(\tau) \cong -4m_s^2 \left(1 + \frac{2|\eta|}{3} \right)\, J_k(\tau)
\end{equation}
To prove this notice that in the limit $-k\tau \rightarrow 0$ the function $\chi$
becomes (see appendix B for more details)
\begin{equation}
\label{large_lim}
  \chi_k(\tau) \rightarrow \frac{H_0}{\sqrt{2 k^3}}\left(-k\tau\right)^{\eta}
\end{equation}
so that, from (\ref{Jk}), we have
\begin{equation}
\label{Jk_zero}
  J_k(\tau) \rightarrow  -p(p-1)\frac{H_0}{\sqrt{2 k^3}}\left(k H_0\right)^{\eta} (-\tau)^{2\eta}
\end{equation}
Now, notice that the operator $\Box_k \rightarrow -H_0^2(\tau^2 \partial_\tau^2 -2\tau\partial_\tau)$ in the same limit.  Acting with this operator on (\ref{Jk_zero})
we have $\Box_k J_k(\tau) = -6|\eta|H_0^2 (1 + 2|\eta|/3) J_k(\tau)$.  Using the definition of the slow roll parameter, $\eta = -2m_s^2 / (3H_0^2)$, we trivially 
recover the result (\ref{Jk_evalue}).

Given the result (\ref{Jk_evalue}) and the method of subsection \ref{J_efunctions_sec} it is easy to compute the correction term on the right-hand-side of (\ref{large_corrections}):
\begin{equation}
  G\left[J_k(\tau)\right] \cong \frac{1}{F\left[-4m_s^2\left(1 + \frac{2|\eta|}{3}\right)\right]} J_k(\tau) = \frac{1}{p^{2(1 + 2|\eta|/3)} - p} J_k(\tau)
\end{equation}
(See equation \ref{J_efunction_G}.)
Thus, we conclude that the solutions of (\ref{p_pert}) take the form
\begin{equation}
\label{phi_large}
  \phi_k(\tau) \cong \chi_k(\tau) \left[ 1 -  \frac{(p-1)}{p^{1+ 4|\eta|/3}-1} u(\tau)  + \mathcal{O}(u^2(\tau)) \right]
\end{equation}
in the limit $-k\tau \rightarrow 0$.  We see that on large scales and to leading order in $u$ the slow motion of $\phi_0$ yields a tiny almost-constant correction to the solutions
obtained in \cite{pi}.  Significantly, we see that slow motion of the background does \emph{not} lead to any super-horizon evolution for the inflaton perturbations (at least at the linearized level
and to first order in $u$), as one would expect in the absence of entropy perturbations \cite{conserved}.

\subsection{Corrections to the $p$-adic Mode Functions}

Now we compute the right-hand-side of (\ref{large_corrections}) without taking the large-scale limit.  To do so, we first decompose the ``source'' term $J_k(\tau)$ - eqn.\ (\ref{Jk}) - 
into a discrete sum of eigenfunctions of $\Box_k$.\footnote{As mentioned previously and discussed in appendix D, such an expansion is not, in general, possible.  Our success here relies on
the particular form of the source term defined by (\ref{Jk}).}  The technical details are discussed in appendix C.  The result is (see equation (\ref{J_result_app}))
\begin{eqnarray}
  J_k(\tau) &=& -p(p-1) \left(\frac{H_0}{k}\right)^{\eta} \left[ \sum_{n=0}^{\infty} \alpha_n^{(1)}\, \frac{H_0 \sqrt{\pi}}{2k^{3/2}}(-k\tau)^{3/2} J_{3/2+2n}(-k\tau)    \right. \nonumber \\
   && \left. + \sum_{n=0}^{\infty} \alpha_n^{(2)} \, \frac{H_0 \sqrt{\pi}}{2k^{3/2}}(-k\tau)^{3/2} J_{-3/2+2\eta+2n}(-k\tau) \right]  \label{J_result}
\end{eqnarray}
where the coefficients $\alpha_n^{(i)}$ depend only on $\eta$ and are explicitly defined by (\ref{alpha1},\ref{alpha2}).  Each member of the sum (\ref{J_result}) is an eigenfunction of $\Box_k$,
as can be seen explicitly from equation (\ref{gen_evalue}).  To compute the right-hand-side of (\ref{large_corrections}) 
we use formula (\ref{J_efunction_G_sum}).  The result is again a sum of the form (\ref{J_result}), only the coefficients have changed:
\begin{eqnarray}
  G\left[J_k(\tau)\right] &=& -p(p-1) \left(\frac{H_0}{k}\right)^{\eta} \left[ \sum_{n=0}^{\infty} \beta_n^{(1)}\, \frac{H_0 \sqrt{\pi}}{2k^{3/2}}(-k\tau)^{3/2} J_{3/2+2n}(-k\tau)    \right. \nonumber \\
   && \left. + \sum_{n=0}^{\infty} \beta_n^{(2)} \, \frac{H_0 \sqrt{\pi}}{2k^{3/2}}(-k\tau)^{3/2} J_{-3/2+2\eta+2n}(-k\tau) \right]  \label{J_G_result}
\end{eqnarray}
(See appendix C for more details.)  The coefficients $\beta^{(1)}_n$, $\beta^{(2)}_n$ in the expansion (\ref{J_G_result}) are defined explicitly in (\ref{beta1},\ref{beta2}).

We now plug the result (\ref{J_G_result}) into (\ref{large_corrections}) and cast the leading order mode functions in the form
\begin{equation}
\label{phi_result}
  \phi_k(\tau) = \chi_k(\tau)\left[ 1 + u(\tau)\Delta(-k\tau) \right]
\end{equation}
In (\ref{phi_result}) we have defined the function $\Delta(-k\tau) \equiv G\left[J_k(\tau)\right]  / \left[u(\tau) \chi_k(\tau) \right]$ which represents the coefficient of the $\mathcal{O}(u)$ contribution to 
fractional difference between $\phi_k(\tau)$ and $\chi_k(\tau)$.  In figure \ref{D1fig} we plot the modulus $|\Delta(x)|$ on large scales $x < 1$.  For illustration we take $\eta = -0.02$ and $p = 5$.  In this figure we 
have retained up to $n = 25$ in the summation (\ref{J_G_result}), however, keeping only up to $n=1$ the curve would have been almost indistinguishable.  We have verified both analytically and numerically that
\begin{equation}
  \lim_{x\rightarrow 0} \Delta(x) = -\frac{p-1}{p^{1+4|\eta|/3} - 1}
\end{equation}
in agreement with (\ref{phi_large}).

In figure \ref{D2fig} we plot $|\Delta(x)|$ for a larger range of $x$ in order to show the large scale behavior of the correction.  Again we take $\eta = -0.02$ and $p = 5$ for illustration.  
Here we see that at large $x$ the function $\Delta(x)$ undergoes oscillations of amplitude $p$.  Notice that even though $u \ll 1$ it may still happen that $u p =\mathcal{O}(1)$ before the end of 
inflation, provided $p$ is sufficiently large.  In that case our result (\ref{phi_result}) for $\phi_k(\tau)$ cases to be reliable on very small scales because higher order terms 
(such as the second and third terms on the second line of (\ref{phi_i})) become important.  This should not be taken as evidence that $\phi_k(\tau)$ differs significantly from $\chi_k(\tau)$ on small scales when $p \gg1$.
It is easy to see that this is not so.   For time intervals $\Delta t \ll H_0^{-1}$ one can treat $u(t)$ as a constant.  
Denoting this constant value by $u_c$ we see from (\ref{p_pert}) that $\phi$ obeys the equation
\[
 \left[ F(\Box_k) + p(p-1)u_c \right] \phi_k(\tau) = 0
\]
The solutions $\phi_k(\tau)$ are given by eigenfunctions of $\Box_k$ where the eigenvalues are given by the zeroes of $F(-\omega^2) + p(p-1)u_c$.  Recall that $\chi_k(\tau)$ is an eigenfunction
of $\Box_k$ with eigenvalue given by the zeroes of $F(-\omega^2)$.  We conclude that on small scales the solutions $\phi_k(\tau)$ of (\ref{p_pert}) differs from $\chi_k(\tau)$ only by a tiny correction to the effective
mass.



\DOUBLEFIGURE[ht]{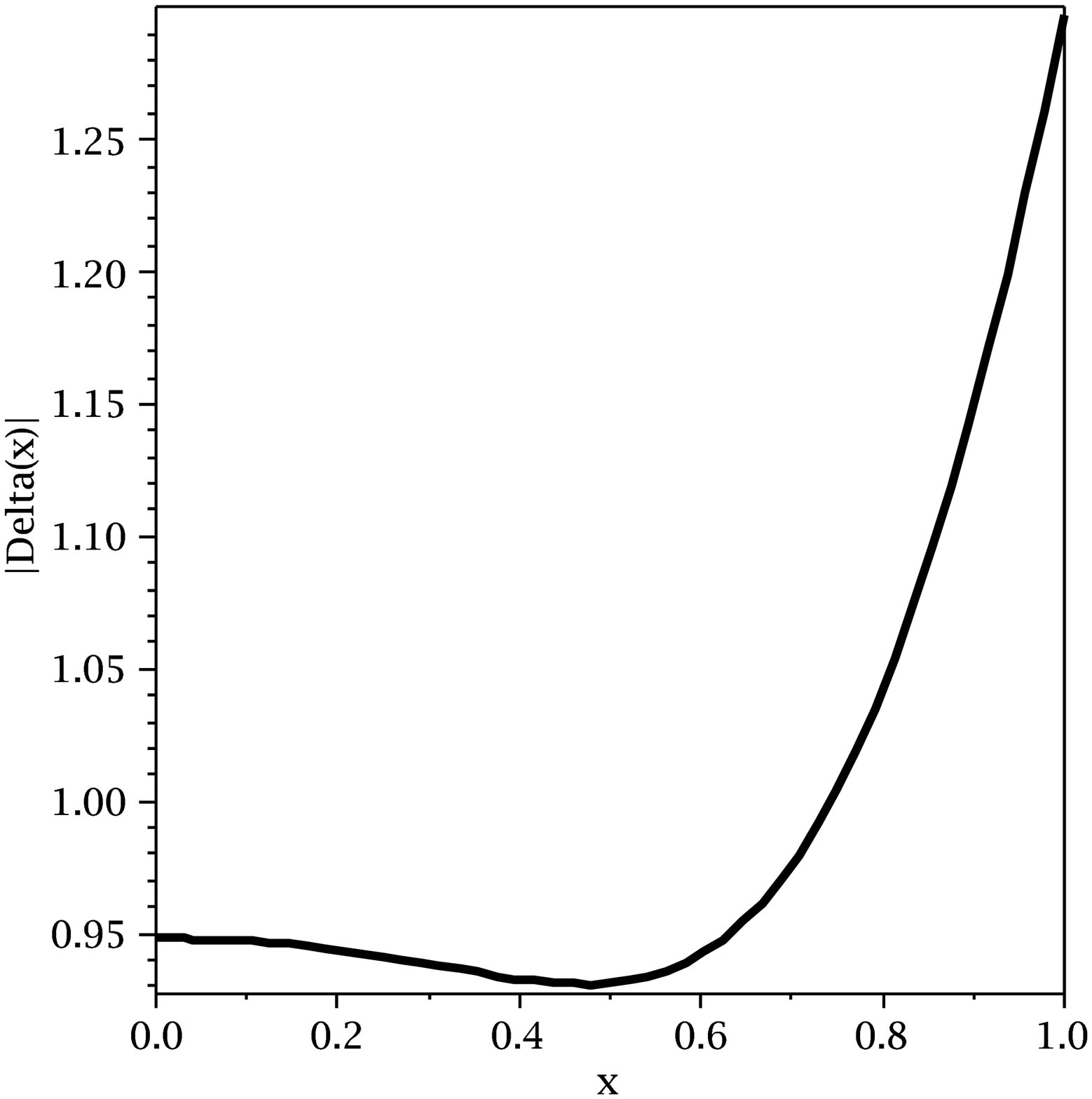,width=2.8in,height=2.8in}{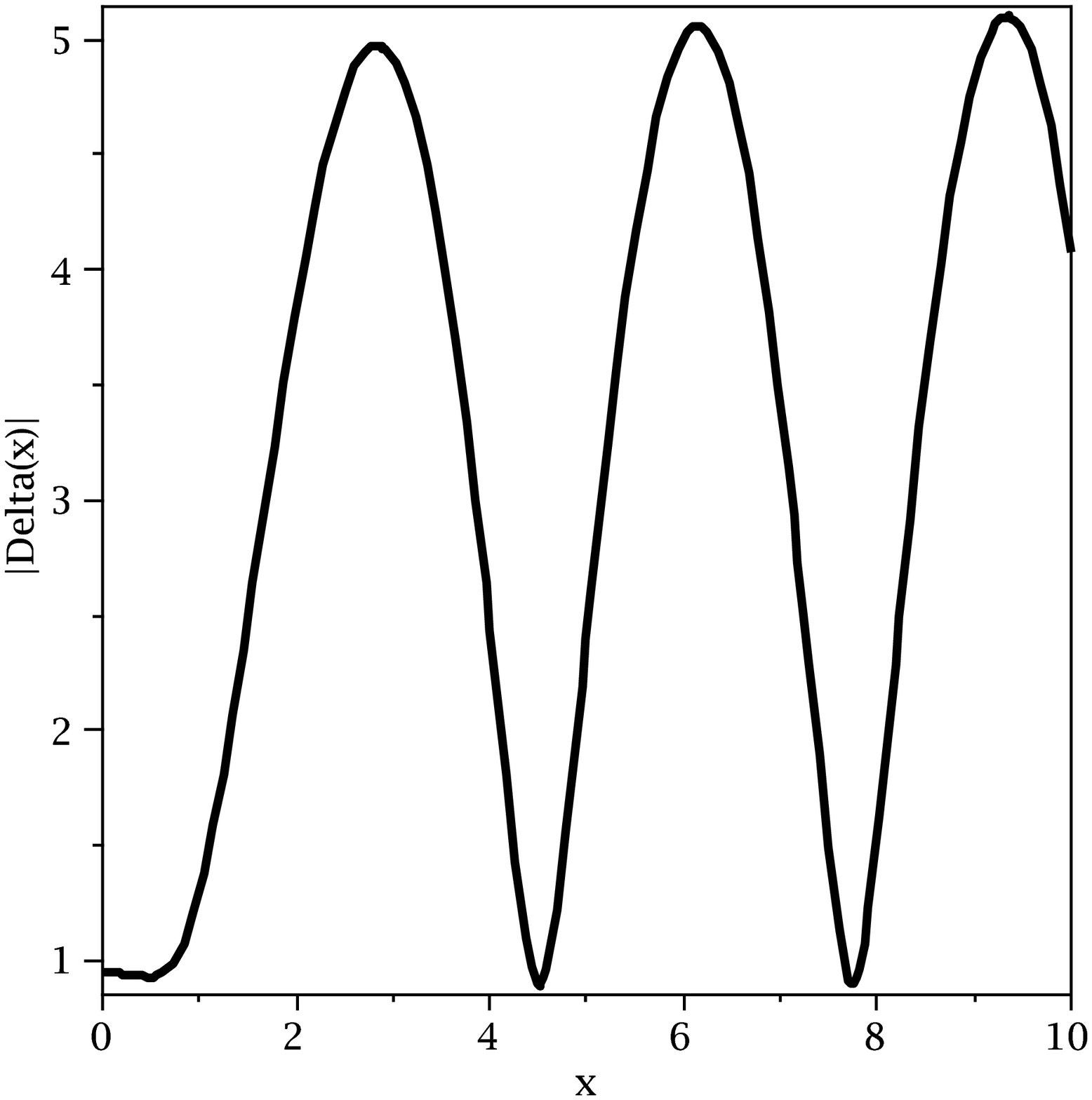,width=2.8in,height=2.8in}{$|\Delta(x)|$ versus $x$ on small scales $x = -k\tau < 1$.  Parameters were chosen as $\eta = -0.02$, $p=5$ for illustration.\label{D1fig}}{$|\Delta(x)|$ versus $x$, showing large scale behaviour.  Parameters were chosen as $\eta = -0.02$, $p=5$ for illustration. \label{D2fig}}

\section{Conclusions and Future Directions}
\label{concl_sec}

In this paper we have developed the theory of a class of variable coefficient equations of infinite order which arise when studying perturbations about a time-dependent
background.  We have shown that solutions can be efficiently computed in terms of the eigenfunctions of the d'Alembertian and the inverse operator associated with the kinetic function $F(\Box)$.
We have illustrated our method for the case of $p$-adic inflation by computing the leading order corrections to the inflaton perturbations derived in \cite{pi}.  We have shown explicitly that 
these are small and do not lead to any super-horizon evolution.

It is worth emphasizing that many of the techniques which we have developed here - in particular the method of successive substitution in section \ref{subst_sec} and the construction
of the nonlocal Green function in section \ref{inverse_sec} - are quite general.  We believe that even ignoring the main themes of this paper such techniques constitute a valuable contribution
to the understanding of infinite order differential equations.

We conclude by discussing some further possible applications of our formalism.  One motivation for this work was to develop the tools necessary for a fully rigorous and systematic
approach to nonlocal cosmological perturbation theory.  In the most general scenario the solution of the perturbed scalar field equation will proceed very much analogously to 
the calculation in section \ref{app_sec}, only the  de Sitter space d'Alembertian is replaced by a perturbed d'Alembertian which includes also metric inhomogeneities.  This replacement significantly complicates the 
construction of the eigenfunctions and also of the resolvent generatrix.  However, no new conceptual obstacle is involved since our formalism does not make any specific assumptions about the detailed form of the metric.  The correlation 
functions of the curvature perturbation may then be computed using the Seery et al.\ formalism for working directly with the field equations \cite{seery}.

Of course, the applications of our formalism are not limited to nonlocal inflation.  This approach could also be used to study perturbations about rolling tachyon solutions in string field theory,
such as the Hellerman and Schnabl \cite{schnabl} solution.  We expect also that our approach could be straightforwardly generalized to study nonlocal dynamics in theories where $F(z)$ has poles
or branch cuts.  Yet another potential application of our method is to the construction of more general nonlocal inflationary background solutions, for example large
field inflation models which might give rise to observable gravitational wave signatures.  We intend to return to these applications in future work.

\section*{Acknowledgments}

This work was supported in part by NSERC. We are grateful to L.\ Boyle for helpful conversations and for providing us with his unpublished notes concerning the Hermiticity
of the d'Alembertian in FRW space-time.  We would like to also thank G.\ Calcagni, D.\ Mulryne, D.\ Seery and R.\ Woodard for interesting discussions, comments and correspondence.

\renewcommand{\theequation}{A-\arabic{equation}}
\setcounter{equation}{0}  

\section*{APPENDIX A: Review of $p$-adic Inflation Background Solutions}

In this appendix we review the slowly rolling solutions obtained in \cite{pi} 
for $p$-adic inflation.  Consider $p$-adic string theory coupled to Einstein-Hilbert gravity:
\begin{equation}
\label{S_app}
  S = \int d^4 x \sqrt{-g}\left[ \frac{M_p^2}{2}R + \mathcal{L}_p   \right]
\end{equation}
where
\begin{equation}
\label{Lp}
  \mathcal{L}_p = \frac{m_s^4}{g_p^2}\left[-\frac{1}{2}\phi p^{-\Box / (2m_s^2)}\phi + \frac{1}{p+1}\phi^{p+1}\right]
\end{equation}
In (\ref{Lp}) $m_s = (\alpha')^{-1/2}$ is the string mass and we have defined
\begin{equation}
  g_p^2  = g_s^2 \frac{p-1}{p^2}
\end{equation}
with $g_s$ the open string coupling constant.  The Lagrangian (\ref{Lp}) is derived for $p$ a prime number, however, the theory can analytically continued to any integer value.

In \cite{pi} inflationary solutions $\phi_0(t)$, $a(t)$ of the theory (\ref{S_app}) we constructed by employing an expansion in powers of $u \equiv e^{\lambda t}$
\begin{eqnarray}
  \phi_0(t) &=& 1 - \sum_{r=1}^{\infty} \phi_r e^{r\lambda t} \label{phi0} \\
  H(t) &=& H_0 - \sum_{r=1}^{\infty} H_r e^{r\lambda t} \label{H0}
\end{eqnarray}
where, of course, $H = \dot{a} / a = \partial_t a / a$.  We have chosen the parametrization of the solutions such that at $t \rightarrow -\infty$ the field $\phi$ starts from the unstable 
maximum of its potential, $\phi = 1$, and the universe undergoes a de Sitter expansion with Hubble constant $H_0$.  As $t$ increases, the corrections terms $e^{r\lambda t}$ become more important and the field
rolls towards the true vacuum $\phi = 0$.  The origin of time is chosen arbitrarily which, of course, has no impact on any physical observable.

The solutions obtained in \cite{pi}, up to order $e^{2\lambda t}$, are
\begin{eqnarray}
  \phi_0 &\cong& 1 - e^{|\eta|H_0 t} + \frac{1}{2}\frac{p-1}{p^{1 + 2|\eta|/3} - 1}\,e^{2|\eta|H_0 t}\label{phi0_soln} \\
  H &\cong& H_0 \left[ 1  - \frac{p \ln p}{2}\,\frac{p+1}{p-1} \,e^{2|\eta|H_0 t}\right] \label{H0_soln}
\end{eqnarray}
In writing (\ref{phi0_soln}-\ref{H0_soln}) we have defined the background Hubble scale
\begin{equation}
\label{false}
  H_0^2 = \frac{m_s^4}{6 M_p^2}\frac{p-1}{g_p^2(p+1)}
\end{equation}
and the slow roll parameter
\begin{equation}
\label{eta}
  \eta = -\frac{2 m_s^2}{3H_0^2}
\end{equation}
which satisfies $|\eta| \ll 1$.  Notice that the small parameter $u(t)$ can be written in terms of the scale factor $a(t)$ as
\begin{equation}
\label{u_to_a}
  u(t) = e^{|\eta|H_0 t} = a(t)^{|\eta|}
\end{equation}

To leading order in $u$ we can consistently treat the background expansion as pure de Sitter space while still working to nontrivial order in the $\eta$ slow
roll parameter.  The consistency of this approach is due to the fact that there is a large hierarchy between the $\eta$ slow roll parameter (eqn.\ \ref{eta})
which controls the size of the inflaton (effective) mass and the $\epsilon$ parameter ($\epsilon = -\dot{H} / H^2$) which controls the departures of the
geometry from de Sitter.  In the same approximation we can write the quantity $m^2(t)$ (\ref{pm}) for the $p$-adic perturbation equation (\ref{p_pert}) as
\begin{equation}
\label{mp_approx}
  m^2(t) \cong -p(p-1) u(t) = -p(p-1)e^{|\eta|H_0 t}
\end{equation}

\renewcommand{\theequation}{B-\arabic{equation}}
\setcounter{equation}{0}  

\section*{APPENDIX B: Construction of $\chi$ for $p$-adic Inflation}

In this appendix we consider the solutions of equation (\ref{chi}) for the case of $p$-adic inflation.  In this case the
kinetic function is given by (\ref{pF}) and hence the equation under consideration is
\begin{equation}
\label{chi_app}
 \left[ p^{-\Box / (2 m_s^2)} - p \right] \chi = 0
\end{equation}
This equation belongs to the class considered in section \ref{const_sec}.  The solutions $\chi_n$ of (\ref{chi_app}) are eigenfunctions of $\Box$
\begin{equation}
\label{chi_n_app}
  \Box \chi_n = -\omega_n^2 \chi_n
\end{equation}
The eigenvalues solve the transcendental equation $F(-\omega^2_n) = 0$ and are given explicitly by \cite{niky}
\begin{equation}
\label{omega_n}
  \omega_n^2 = \left[ 2 \pm \frac{4\pi i n}{\ln p} \right] m_s^2
\end{equation}
with $n = 0, 1, \cdots$  The $n=0$ mode is the usual tachyon with effective mass $-\omega^2 = -2m_s^2$.  The infinity of $n > 0$ states have complex mass-squared and are ghost-like (contributing
negative kinetic energy to the Hamiltonian); see \cite{mulryne},\cite{null} and \cite{pais}.  These states are, presumably, artifacts that would not be present in the full string theory.  Here we simply omit the ghost modes 
and focus our attention on the tachyon.  Since we only consider $n=0$ we will drop the subscript $n$ on the eigenfunctions.

The $n=0$ solution of (\ref{chi_n_app}) in a de Sitter geometry is well known
\begin{eqnarray}
  \chi(t,{\bf x}) &=& \int \frac{d^3 k}{(2\pi)^{3/2}}\left[ a_{{\bf k}} e^{i {\bf k}\cdot {\bf x}} \chi_{\bf k}(t) + \mathrm{h.c.} \right] \label{chi1}\\
  \chi_{\bf k}(t) &=& \frac{e^{i\delta}}{2}\sqrt{\frac{\pi}{a^3 H_0}}H_{\nu}^{(1)}\left(\frac{k}{a H_0}\right) \label{chi2}
\end{eqnarray}
where $a_k$, $a_k^{\dagger}$ are annihilation/creation operators, $\chi_k$ are c-number valued mode functions, $\mathrm{h.c.}$ denotes the Hermitian conjugate of the preceding term and $a = e^{H_0 t}$
is the scale factor.  In (\ref{chi2}) the real-valued constant $\delta$ is an irrelevant phase the order of the Hankel functions is 
\begin{eqnarray}
  \nu  &=& \sqrt{\frac{9}{4} + \frac{\omega_{0}^2}{H_0^2}} \nonumber \\
       &\cong& \frac{3}{2} - \eta + \mathcal{O}(\eta^2)
\end{eqnarray}
to leading order in the $\eta$ slow roll parameter.

It is convenient to introduce conformal time $\tau$ related to cosmic time $t$ as $a d\tau = dt$.  In terms of conformal time the scale factor is
\begin{equation}
  a(\tau) = -\frac{1}{H_0 \tau}
\end{equation}
so that $\tau$ runs from $-\infty$ to $0$ as $a$ goes from $0$ to $+\infty$.  Notice that $k / (aH_0) = -k\tau$.  The small parameter $u$ in (\ref{u_to_a}) can be written as
\begin{equation}
  u(\tau) = (-H_0\tau)^{-|\eta|}
\end{equation}

Small (sub-horizon) scales corresponds to $-k\tau \gg 1$.  By construction, the solutions (\ref{chi2}) have the following small scale asymptotics
\begin{equation}
\label{chi_small}
  \chi_k \rightarrow \frac{1}{a}\frac{e^{-ik\tau}}{\sqrt{2k}} \hspace{5mm}\mathrm{for}\hspace{5mm}-k\tau \rightarrow \infty
\end{equation}
This normalization corresponds to the usual Bunch-Davies vacuum choice.  Large (super-horizon) scales corresponds to $-k\tau \ll 1$.  The 
solutions (\ref{chi2}) have the following large scale asymptotics
\begin{equation}
\label{chi_large}
  \chi_k \rightarrow \frac{H_0}{\sqrt{2 k^3}}\left(-k\tau\right)^{\eta}\hspace{5mm}\mathrm{for}\hspace{5mm}-k\tau \rightarrow 0
\end{equation}
The factor $k^{-3/2}$ corresponds to the exactly scale invariant part while the factor $(-k\tau)^{\eta}$ give the slight departure from scale invariance with
spectral index $n_s = 1 + 2\eta$.

\renewcommand{\theequation}{C-\arabic{equation}}
\setcounter{equation}{0}  

\section*{APPENDIX C: Eigenfunction Decomposition of $J_k(\tau)$}

In this appendix we would like to show that the quantity $J_k(\tau)$, defined by
\begin{equation}
\label{Jk_app}
  J_k(x) = A_k (-k\tau)^{3/2 + \eta}H_{3/2-\eta}^{(1)}(-k\tau)
\end{equation}
where
\begin{equation}
  A_k = -p(p-1)\frac{H_0 \sqrt{\pi}}{2 k^{3/2}}\left(\frac{H_0}{k}\right)^{\eta}
\end{equation}
(see eqn.\ \ref{Jk}) can be expanded into a discrete sum of eigenfunctions of the operator $\Box_k$ in de Sitter space, defined by (\ref{box2}).

Before proceeding, let us briefly review some facts about the eigenfunctions of $\Box_k$.  It is straightforward to verify the identity
\begin{equation}
\label{gen_evalue}
  \Box_k \left[ (-k\tau)^{3/2} C_{\alpha}(-k\tau)  \right] = \left(\frac{9}{4} - \alpha^2 \right)\,H_0^2 \,\left[ (-k\tau)^{3/2} C_{\alpha}(-k\tau)  \right]
\end{equation}
which is valid for $C_{\alpha}$ any solution of Bessel's equation with order $\alpha$.  In particular, equation (\ref{gen_evalue}) is valid for $C_{\alpha} = H_{\alpha}^{(1)}, H_{\alpha}^{(2)}, J_{\alpha}, Y_{\alpha}$
(or any linear combination thereof).  

As a first step to decompose $J_k(\tau)$ into a sum of eigenfunctions of $\Box_k$, then, we should write $x^{\eta}H_{3/2-\eta}^{(1)}(x)$ as sum of terms of the form $C_{\alpha}(x)$
(here we have denoted $x\equiv -k\tau$).  To this end we write the Hankel function in terms of Bessel functions of the first kind as
\begin{eqnarray}
  x^{\eta} H_{3/2-\eta}^{(1)}(x) &=& x^{\eta}\left[ J_{3/2-\eta}(x) + i Y_{3/2-\eta}(x)\right] \nonumber \\
                        &=& \left(1 + i \cot\left[(3/2-\eta)\pi\right]\right)\, x^{\eta} J_{3/2-\eta}(x) \nonumber \\
                        && - \frac{i}{\sin\left[(3/2-\eta)\pi\right]}\, x^{\eta} J_{-3/2+\eta}(x) \label{hankel}
\end{eqnarray}
(this follows from the definitions of $H_{\nu}^{(1)}$ and $Y_{\nu}$).  Next, we employ the identity
\begin{eqnarray}
  \left(\frac{x}{2}\right)^{\mu-\nu}J_{\nu}(x) &=& \sum_{n=0}^{\infty} c(n;\mu,\nu) J_{\mu + 2n}(x) \label{J_decomp} \\
  c(n;\mu,\nu) &\equiv& \frac{\mu + 2n}{n!}\frac{\Gamma(\mu+n)\Gamma(\nu-\mu+1)}{\Gamma(\nu-\mu+1-n)\Gamma(\nu+1 + n)} \label{c}
\end{eqnarray}
which is derived on page 139 of \cite{bessel}.  Using (\ref{J_decomp}) we can we-write the quantities $x^{\eta}J_{3/2-\eta}(x)$ and $ x^{\eta} J_{-3/2+\eta}(x)$ 
appearing on the second and third lines of (\ref{hankel}) in terms of Bessel functions of the first kind.  The result is:
\begin{eqnarray}
  && x^{3/2 + \eta} H_{3/2-\eta}^{(1)}(x) = 2^{\eta}\,\left[1 + i\cot\left[\left(3/2-\eta\right)\pi\right]\right]\sum_{n=0}^{\infty}c(n;3/2,3/2-\eta)\, x^{3/2} J_{3/2 + 2n}(x) \nonumber \\
  &&             - \frac{2^{\eta}\,i}{\sin\left[\left(3/2-\eta\right)\pi\right]}\sum_{n=0}^{\infty}c(n;-3/2+2\eta,-3/2+\eta)\, x^{3/2} J_{-3/2 + 2\eta + 2n}(x) \label{identity}
\end{eqnarray}
Using the identity (\ref{identity}) we can finally decompose (\ref{Jk_app}) into a sum of eigenfunctions of $\Box_k$ as follows:
\begin{eqnarray}
  J_k(\tau) &=& -p(p-1) \left(\frac{H_0}{k}\right)^{\eta} \left[ \sum_{n=0}^{\infty} \alpha_n^{(1)}\, \frac{H_0 \sqrt{\pi}}{2k^{3/2}}(-k\tau)^{3/2} J_{3/2+2n}(-k\tau)    \right. \nonumber \\
   && \left. + \sum_{n=0}^{\infty} \alpha_n^{(2)} \, \frac{H_0 \sqrt{\pi}}{2k^{3/2}}(-k\tau)^{3/2} J_{-3/2+2\eta+2n}(-k\tau) \right]  \label{J_result_app}
\end{eqnarray}
where the coefficients $\alpha_n^{(i)}$ depend only on $\eta$:
\begin{eqnarray}
  \alpha_n^{(1)} &=& 2^{\eta}\left(1 + i\cot\left[\left(3/2-\eta\right)\pi\right]\right)\, c(n;3/2,3/2-\eta) \label{alpha1} \\
  \alpha_n^{(2)} &=& \frac{ -2^{\eta}\, i}{\sin\left[\left(3/2-\eta\right)\pi\right]}\, c(n;-3/2+2\eta,-3/2+\eta) \label{alpha2}
\end{eqnarray}
Equation (\ref{J_result_app}) is the main result of this appendix.  The terms in the summation (\ref{J_result}) should be compared to the de Sitter modes (\ref{mode}).
This shows that each term in the expansion (\ref{J_result_app}) behaves very much like the mode function for a massive field in de Sitter space.

If we introduce the notation
\begin{eqnarray}
  \psi_n^{(1)} &=& \frac{H_0 \sqrt{\pi}}{2k^{3/2}}(-k\tau)^{3/2} J_{3/2+2n}(-k\tau) \label{psi1} \\
  \psi_n^{(2)} &=& \frac{H_0 \sqrt{\pi}}{2k^{3/2}}(-k\tau)^{3/2} J_{-3/2+2\eta+2n}(-k\tau) \label{psi2}
\end{eqnarray}
for the eigenfunctions appearing in the summation (\ref{J_result_app}) then, from (\ref{gen_evalue}), we have
the identities
\begin{eqnarray}
  \Box_k \psi_n^{(1)} &=& \left[-4n^2 - 6n \right]H_0^2\, \psi_n^{(1)} \\
  \Box_k \psi_n^{(2)} &=& \left[-4(n+\eta)^2 + 6(n + \eta) \right]H_0^2\, \psi_n^{(2)} 
\end{eqnarray}
Using equation (\ref{J_efunction_G}) we can easily compute $G\left[\psi_k^{(i)}(\tau)\right]$.  We find
\begin{eqnarray}
  G\left[ \psi_n^{(1)}\right] &=&  \frac{1}{F\left[\left(-4n^2 - 6n \right)H_0^2\right]}\, \psi_n^{(1)} \\
  G\left[ \psi_n^{(2)}\right] &=& \frac{1}{F\left[\left(-4(n+\eta)^2 + 6(n + \eta) \right)H_0^2\right]}\, \psi_n^{(2)} 
\end{eqnarray}
where $F(z)$ is defined by (\ref{pF}).  Since the resolvent $G$ is linear we have
\begin{eqnarray}
  G\left[J_k(\tau)\right] &=& -p(p-1) \left(\frac{H_0}{k}\right)^{\eta} \left[ \sum_{n=0}^{\infty} \beta_n^{(1)}\, \frac{H_0 \sqrt{\pi}}{2k^{3/2}}(-k\tau)^{3/2} J_{3/2+2n}(-k\tau)    \right. \nonumber \\
   && \left. + \sum_{n=0}^{\infty} \beta_n^{(2)} \, \frac{H_0 \sqrt{\pi}}{2k^{3/2}}(-k\tau)^{3/2} J_{-3/2+2\eta+2n}(-k\tau) \right]  \label{J_G_result_app}
\end{eqnarray}
where the coefficients are
\begin{eqnarray}
 \beta_n^{(1)} &=& \frac{\alpha_n^{(1)}}{p^{\frac{2}{3|\eta|}\left[2n^2 + 3n\right]} - p} \label{beta1} \\
  \beta_n^{(2)} &=& \frac{\alpha_n^{(2)}}{p^{\frac{1}{3|\eta|}\left[2(n+\eta)^2-3(n+\eta)\right]} - p} \label{beta2}
\end{eqnarray}
In deriving (\ref{beta1},\ref{beta2}) we have used the fact that $\eta = -2m_s^2 / (3H_0^2)$.


\renewcommand{\theequation}{D-\arabic{equation}}
\setcounter{equation}{0}  

\section*{APPENDIX D: On the Non-Self-Adjointness of the d'Alembertian}

In this appendix we review some basic facts about Sturm-Liouville theory to demonstrate that the operator $\Box$ fails to be self-adjoint
in a de Sitter geometry when the function space contains the standard Bunch-Davies mode functions.  We are interested in the eigenvalue equation
\begin{equation}
\label{app_efunction_box}
  \Box v_m(\tau,{\bf x}) = m^2 v_m(\tau,{\bf x})
\end{equation}
In a de Sitter geometry where $\Box$ is given by (\ref{box1}).  The equation for the Fourier modes can be cast in the standard Sturm-Liouville form
as
\begin{equation}
\label{u_efunction}
  L\left[v_{k,m}(\tau)\right] + m^2 a^4(\tau)\, v_{k,m}(\tau) = 0
\end{equation}
where the Sturm-Liouville operator is
\begin{equation}
\label{sl_op}
  L  \equiv \partial_\tau\left[ a^2(\tau)\partial_\tau \right] + k^2 a^2(\tau)
\end{equation}
the eigenvalue is $\lambda  = m^2$ and the weight is $a^4(\tau)$ (with $a(\tau)=-1/(H_0\tau)$).  

Although (\ref{sl_op}) is in Sturm-Liouville form, it may fail to be self-adjoint if we impose physically interesting boundary 
conditions on the eigenfunctions.  To see explicitly how this failure occurs, let us suppose that our function space contains the functions $f_k(\tau)$, $g_k(\tau)$ defined by:
\begin{eqnarray}
  f_{k}(\tau) &=& \frac{H_0 \sqrt{\pi}}{2k^{3/2}}(-k\tau)^{3/2} H_{\nu_f}^{(1)}(-k\tau) \label{f}\\
  g_{k}(\tau) &=& \frac{H_0 \sqrt{\pi}}{2k^{3/2}}(-k\tau)^{3/2} H_{\nu_g}^{(1)}(-k\tau) \label{g}
\end{eqnarray}
where 
\begin{equation}
  \nu_{f} = \sqrt{\frac{9}{4} - \frac{m_{f}^2}{H_0^2}}, \hspace{5mm}\nu_{g} = \sqrt{\frac{9}{4} - \frac{m_{g}^2}{H_0^2}}
\end{equation}
The functions $f_k(\tau)$, $g_k(\tau)$ correspond to de Sitter space mode functions normalized according to the Bunch-Davies prescription.  For simplicity we assume that $0 < m_{f}, m_{g} < H$,
however, our analysis does not rely on this assumption in any crucial way.

Working on the interval $\tau \in \left(\tau_1,\tau_2\right)$  the standard theory gives the identity
\begin{eqnarray}
  && \int_{\tau_1}^{\tau_2}d\tau f_{k}^{\star}(\tau) L \left[g_{k}(\tau)\right] - \int_{\tau_1}^{\tau_2}d\tau g_{k}(\tau) L \left[f_{k}(\tau)^{\star}\right] \nonumber \\
  && = \left[ a^2(\tau)\left( f_{k}(\tau)^{\star}\partial_\tau g_{k}(\tau) - g_{k}(\tau)\partial_\tau f_{k}^{\star}(\tau)  \right)\right]_{\tau=\tau_1}^{\tau_2} \label{hermitian}
\end{eqnarray}
Let us now show explicitly that term on the second line of (\ref{hermitian}) does not vanish when one takes the physically sensible endpoints $\tau_1 = -\infty$, $\tau_2 = 0$. 
 Using the known large- and small-scale asymptotics of the de Sitter space mode functions (see appendix B) it is straightforward to show that
\begin{equation}
\label{thing1}
  \left[ a^2(\tau)\left( f_{k}^{\star}(\tau)\partial_\tau g_{k}(\tau) - g_{k}(\tau)\partial_\tau f_{k}^{\star}(\tau)  \right)\right] \rightarrow -i
\end{equation}
in the limit $\tau \rightarrow -\infty$.  On the other hand, we have
\begin{equation}
\label{thing2}
    \left[ a^2(\tau)\left( f_{k}(\tau)^{\star}\partial_\tau g_{k}(\tau) - g_{k}(\tau)\partial_\tau f_{k}^{\star}(\tau)  \right)\right] \rightarrow \frac{(\nu_g-\nu_f)}{2}\left(-k\tau\right)^{-\nu_f-\nu_g}
\end{equation}
in the limit $-\tau \rightarrow 0$.  Clearly (\ref{thing1}) does not equal (\ref{thing2}) and this completes our proof that $L$ is not self-adjoint.

It is a simple exercise to see that, because
\[
  \int_{-\infty}^{0}d\tau f_{k}^{\star}(\tau) L \left[g_{k}(\tau)\right] \not= \int_{-\infty}^{0}d\tau g_{k}(\tau) L \left[f_{k}^{\star}(\tau)\right]
\]
the usual theorems about orthogonality and completeness of the eigenfunctions $v_{k,m}(\tau)$ on the interval $-\infty < \tau < 0$ do not hold.  Hence, there is no reason to expect that a general source $J_k(\tau)$
in the function space will be expansible in a series of eigenfunctions.  Of course, this analysis does not rule out the possibility that some physically interesting source terms, such as (\ref{Jk}), may happen
to be so expansible (see appendix C).  However, this is a special property of the particular source defined by (\ref{Jk}), rather than a generic property of the eigenfunctions of $\Box$ in de Sitter space.

Note that physically the failure of $L$ to be self-adjoint is easy to understand.  It occurs because the mode functions have very different behaviour on large scale as compared to the small scale asymptotics.
Hence we expect similar behaviour in general FRW space-times.

It is interesting - if tangential to our main line of inquiry - to consider more carefully the reason that $\Box$ is not self-adjoint in de Sitter space.  The choice of Bunch-Davies mode functions is closely tied
to the choice of conformal coordinates defined by
\begin{equation}
\label{planar}
  ds^2 = a^2(\tau)\left[-d\tau^2 + dx_idx^i\right]
\end{equation}
with $-\infty < \tau < 0$ and $a(\tau) = -1/(H_0\tau)$.  But it is well-known that these coordinates do not cover the whole de Sitter space-time \cite{dSreview}.
Let us reconsider this exercise using global coordinates \cite{dSreview} defined by
\begin{equation}
\label{global}
  ds^2 = -d\tilde{t}^2 + 4 \cosh^2(\tilde{t}) dx^i dx_i
\end{equation}
Taking $-\infty < \tilde{t} < +\infty$ these coordinates cover the full de Sitter space-time.  One could reconsider the solutions of the eigenvalue equation $\Box v_m = m^2 v_m$ in this coordinate system,
however, this effort is not necessary.  In the asymptotic regions $\tilde{t} \rightarrow \pm \infty$ the metric (\ref{global}) corresponds to a FRW universe with scale factor $a(\tilde{t}) \sim e^{\pm H_0\tilde{t}}$.  
Hence, we can normalize the eigenfunctions such that
\begin{equation}
\label{global_mode}
  v_{k,m}(\tilde{t}) \sim \frac{e^{i\delta}}{2}\sqrt{\frac{\pi}{a^3 H_0}}H_{\nu}^{(1)}\left(\frac{k}{a H_0}\right)
\end{equation}
asymptotically as $\tilde{t} \rightarrow \pm \infty$ (see eqn.\ \ref{chi2}).  Due to the symmetry under $\tilde{t} \rightarrow -\tilde{t}$ we have same kind of behaviour for the modes in each asymptotic
region (that is freeze-out).   Now there is no inconsistency with imposing periodic boundary conditions on the eigenfunctions at $\tilde{t}_1 = -T$ and $\tilde{t}_2 = +T$.  These boundary conditions render
$\Box$ self-adjoint and one can send $T \rightarrow \infty$ at the end of the calculation.

However, the possibility of using (\ref{global}) (or, indeed, \emph{any} global covering) to render $\Box$ self-adjoint is quite irrelevant for our purposes.  Certainly the use of global coordinates is inappropriate for the
physical problem at hand.  First off, the covering (\ref{global}) contains an unphysical contracting region for $\tilde{t} < 0$.  Moreover, we are not actually interested in real de Sitter space-time but rather inflationary
spaces that merely mimic de Sitter for some number of e-foldings.  Hence, for the physically reasonable choice of coordinates and vacuum the non-self-adjointness of $\Box$ is simply something we need to live
with.

The reader who still finds the non-self-adjointness of $\Box$ disturbing may wish to consider an analogy.  Using the standard Minkowski coordinates $ds^2 = -dt^2 + dx_idx^i$ one has, in Fourier space, oscillatory eigenfunctions 
$e^{-i\omega t}$.  We can ensure that $\Box$ is self-adjoint by imposing periodic boundary conditions at $\pm T$ and sending $T \rightarrow \infty$ at the end of the calculation.  But now consider the 
same exercise using Milne coordinates.  The Milne universe looks like an expanding FRW space-time with scale factor that grows linearly in cosmic time, however, it is quite equivalent to ordinary flat Minkowski
space-time after a coordinate transformation.  In conformal coordinates the Milne metric takes the form
\begin{equation}
\label{milne}
  ds^2 = a^2(\tau) \left[-d\tau^2 + dr^2 + \sinh^2 r \left( d\theta^2 + \sin^2\theta d\phi^2 \right)\right]
\end{equation}
with $-\infty < \tau < \infty$ and $a(\tau) = e^{\tau}$.  These coordinates do not cover the full space-time but only the patch $t^2 - x_ix^i > 0$.  The analogue of the Bunch-Davies
modes (\ref{f},\ref{g}) are solutions of the form
\begin{eqnarray}
  f_{\lambda}(\tau) &=& \sqrt{\frac{\pi}{2}}\frac{1}{\sqrt{\sinh(\pi \lambda)}}\, J_{-i\lambda}\left(m_f e^{\tau}\right)  \label{f_milne} \\
  g_{\lambda}(\tau) &=& \sqrt{\frac{\pi}{2}}\frac{1}{\sqrt{\sinh(\pi \lambda)}}\, J_{-i\lambda}\left(m_g e^{\tau}\right)  \label{g_milne}
\end{eqnarray}
corresponding to the conformal vacuum \cite{lev}.  (Here the real number $\lambda$ is the eigenvalue of the spatial Laplacian and plays the role of $k$ in the de Sitter modes (\ref{f},\ref{g}).)  The modes 
(\ref{f_milne},\ref{g_milne}) behave very differently near the endpoints $\tau = \pm \infty$ and we encounter the same difficulty that we had in the de Sitter case.  The boundary conditions implied by the conformal 
vacuum are inconsistent with the self-adjointness of the d'Alembertian in Minkowski space.

\end{document}